\documentclass[conference]{IEEEtran}
\IEEEoverridecommandlockouts
\usepackage{balance}
\usepackage{ulem}
\usepackage[numbers,sort&compress]{natbib}
\usepackage{amsthm,amsmath,amssymb,thmtools}
\usepackage{mathrsfs}
\usepackage{graphicx}
\usepackage{color}
\usepackage{enumitem}
\makeatletter
\newcommand*\bigcdot{\mathpalette\bigcdot@{.5}}
\newcommand*\bigcdot@[2]{\mathbin{\vcenter{\hbox{\scalebox{#2}{$\m@th#1\bullet$}}}}}
\makeatother
\usepackage{epstopdf}
\usepackage{booktabs}
\usepackage{bm}
\normalsize
\usepackage{caption}
\usepackage{subcaption}
\declaretheoremstyle[
  headfont=\normalfont\itshape,
  headpunct=\textup{:},
  bodyfont=\normalfont,
  headindent=1em
]{myremark}
\usepackage[algo2e,ruled,linesnumbered]{algorithm2e}
\makeatletter
\newcommand{\nosemic}{\renewcommand{\@endalgocfline}{\relax}}
\newcommand{\dosemic}{\renewcommand{\@endalgocfline}{\algocf@endline}}
\let\oldnl\nl
\newcommand{\nonl}{\renewcommand{\nl}{\let\nl\oldnl}}
\makeatother
\SetAlCapHSkip{0pt}
\SetKwIF{If}{ElseIf}{Else}{if}{then}{else~if}{else}{end~if}
\usepackage{float}
\usepackage{setspace}

\newlength{\textfloatsepsave} 

\usepackage{algpseudocode}
\usepackage{amsmath}

\begin{document}
\title{Memory-Efficient Split Federated Learning for LLM
	Fine-Tuning on Heterogeneous Mobile Devices
	\author{\IEEEauthorblockN{
			Xiaopei Chen\IEEEauthorrefmark{1}\IEEEauthorrefmark{2},
			Liang Li\IEEEauthorrefmark{2},
			Fei Ji\IEEEauthorrefmark{3}, and
			Wen Wu\IEEEauthorrefmark{2}
		} 
		\IEEEauthorblockA{\IEEEauthorrefmark{1}School of Future Technology, South China University of Technology, Guangzhou, China}
		\IEEEauthorblockA{\IEEEauthorrefmark{2}Frontier Research Center, Peng Cheng Laboratory, Shenzhen, China}
		\IEEEauthorblockA{\IEEEauthorrefmark{3}School of Electronic and Information Engineering, South China University of Technology, Guangzhou, China}
		
		Email:{\{chenxp, lil03, wuw02\}@pcl.ac.cn, eefeiji@scut.edu.cn}\\
	}
	\thanks{\scriptsize{This work was supported in part by NSFC 62201071 and 62192712, in part by the Peng Cheng Laboratory Major Key Project under Grants PCL2023AS1-5 and PCL2021A09-B2, in part by NSFC 62201311, and in part by the Young Elite Scientists Sponsorship Program by CAST under Grant 2023QNRC001. (\textit{Corresponding author: Liang Li.})}}
}

\maketitle

\begin{abstract} In this paper, we propose an edge-assisted split federated learning framework to facilitate large language model (LLM) fine-tuning on heterogeneous mobile devices while alleviating memory pressures on both mobile devices and the edge server. Specifically, mobile devices perform low-rank adaptation (LoRA) fine-tuning on only a subset of lower layers of the pre-trained LLM, tailored to their individual capacities. On the server, a full LLM is maintained, and the corresponding LoRA modules are selectively fine-tuned in a sequential manner for each device. To further enhance training efficiency, we propose a server-side training scheduling method that optimizes the processing order of devices for accelerating fine-tuning. Extensive experiments demonstrate that compared to the baselines, our scheme can reduce 79\% memory footprint and 6\% training time while achieving comparable performance.
\end{abstract}

\IEEEpeerreviewmaketitle

\section{Introduction}

\IEEEPARstart {R}{iding} the wave of advancements in Artificial Intelligence (AI), we are witnessing the eruption of Transformer-based large language models (LLMs) such as BERT \cite{devlin2018bert}, GPT \cite{radford2018improving}, and ViT \cite{he2022masked}, which have demonstrated remarkable capabilities across diverse domains including natural language processing (NLP), computer vision (CV), and beyond \cite{9651548, 9749222}. The impressive adaptability of LLMs stems from the pre-training and fine-tuning paradigm, wherein a pre-trained foundation model acquires domain-specific knowledge through fine-tuning, thus enhancing its generalization ability and suitability for specific tasks. To further advance the personalization capabilities of LLMs, there is an increasing emphasis on utilizing local data with privacy preservation for on-device fine-tuning. Achieving efficient fine-tuning at the edge remains a significant challenge due to the limited and heterogeneous resources of mobile devices \cite{10400885}, but the substantial resource demands of LLMs.

To reduce the resource demands of on-device LLM fine-tuning, low-rank adaptation (LoRA) \cite{hu2021lora}, a representative method of parameter-efficient fine-tuning (PEFT), has emerged as a promising solution. LoRA introduces low-rank matrices to adapt pre-trained models to downstream tasks, significantly reducing the number of parameters that need to be updated during fine-tuning. By freezing the majority of the model's parameters and optimizing only a small subset, LoRA reduces computational overhead, making it highly suitable for deployment in resource-constrained environments like mobile devices. This innovation not only facilitates task-specific adaptation but also retains the general knowledge encoded in the pre-trained model, ensuring efficiency and effectiveness.

Building upon the strengths of PEFT techniques like LoRA, federated learning (FL) based LLM fine-tuning \cite{cai2023efficient,wang2024federated,9488839,10666083} has been proposed to further enable the collaborative LLM fine-tuning on distributed privacy data across multiple mobile users. FL enables multiple clients to train a global model collaboratively  without sharing their local data, ensuring enhanced privacy protection. In these FL frameworks, each mobile device with private data holds a pre-trained LLM for local training and then uploads the LoRA parameters to the server for aggregation updates.  The FL fine-tuning raises a concern about the device's memory because resource-limited mobile devices struggle to cope with the original LLM \cite{10574376}. Although PEFT methods can reduce computational overhead, they require storing intermediate activations to compute the gradients of the trainable parameters, which need a significant memory requirement (over 70\% full fine-tuning \cite{liao2024make}). 

Split Federated Learning (SFL) combines the principles of FL and split learning (SL) by splitting the model between the client and server, allowing clients to train part of the model locally while offloading the remaining computation to the server, thus enhancing privacy and reducing client-side memory usage.  Therefore, some recent attempts have begun to explore the SFL-based LLM fine-tuning paradigm.  For instance, in FedBERT \cite{tian2022fedbert}, the server updates the transformer layers for all clients, while the clients train the embedding and header layers locally. Obviously, FedBERT is suitable for  scenarios where all devices have weak computing capacity. In the scenarios of heterogeneous devices, homogeneous splitting makes it difficult to achieve efficient resource exploitation (wasting the resources of the devices and increasing the burden on the server). SplitLoRA \cite{lin2024splitlora} deploys a homogeneous client-side submodel (embedding layer and the same transformer layers) across all clients, and maintains the residual layers of LLM on the server-side for collaborative fine-tuning. After multiple rounds, the client-side LoRA adapters are updated in an aggregated manner. Current SFL-based LLM fine-tuning focus only on homogeneous configurations of client-side submodels. SplitLoRA \cite{lin2024splitlora} can not cope with heterogeneous client-side models because heterogeneous LoRA cannot be aggregated directly.   On the other hand,  FedBERT \cite{tian2022fedbert}  requires the server to maintain and execute multiple server-side submodels, each corresponding to a different client-side submodels. This will put a huge strain on the server in memory and computing resources and thus pose scalability concerns. 

To tackle the above challenges, we propose an SFL framework for memory-efficient fine-tuning over heterogeneous devices. Specifically, we split the input layer and some Transformer layers of the LLM as client-side submodels deployed to heterogeneous devices according to the computation and memory resources of each device. Unlike traditional SFL, we maintain the entire LLM at the server instead of multiple server-side submodels corresponding to multiple clients. After completing the forward computation of the client-side submodel, the forward activations are transmitted to the server to complete the training of the remaining models. Specifically, the submodel completed on the client is skipped, and the remaining model is utilized to update the server-side LoRA. The proposed scheme sequentially updates the server-side LoRA by reusing the server's entire LLM, which reduces the memory and computational pressure on the server. The main contributions are as follows:
\begin{itemize}
	\item We propose a memory-efficient SFL framework for LLM fine-tuning over heterogeneous devices, which explores the efficient utilization of the SFL framework in the heterogeneous device scenario.
	\item We design a scheduling algorithm to integrate the proposed framework to achieve the trade-off between memory and training time.
\end{itemize}
 
The remainder of the paper is organized as follows. Section~II introduces the system model. Section~III presents the memory-efficient SFL framework. Section~IV proposes a training sequence scheduling algorithm. Section~V provides simulation results. Section~VI concludes this work.

\section{System Model}
As shown in Figure 1, we consider a typical two-tier architecture, which consists of an edge server and a set of mobile devices (clients) denoted by $\mathcal{U} = \{1,2,...,u,..., U\}$. The edge server, with more powerful computational and memory capability, is primarily responsible for the model fine-tuning training task and manages model aggregation and split.   These mobile devices are heterogeneous, with different computing capabilities and memory. Each client $u$ has a local dataset $D_u$ for fine-tuning and the local datasets of the clients are non-independent and identically distributed (Non-IID). The edge server and the mobile devices collaboratively fine-tune a transformer-based LLM by the LoRA method \cite{hu2021lora} for a specific downstream task.
\begin{figure}
	\centering 
	\centering
	\includegraphics*[width=0.49\textwidth]{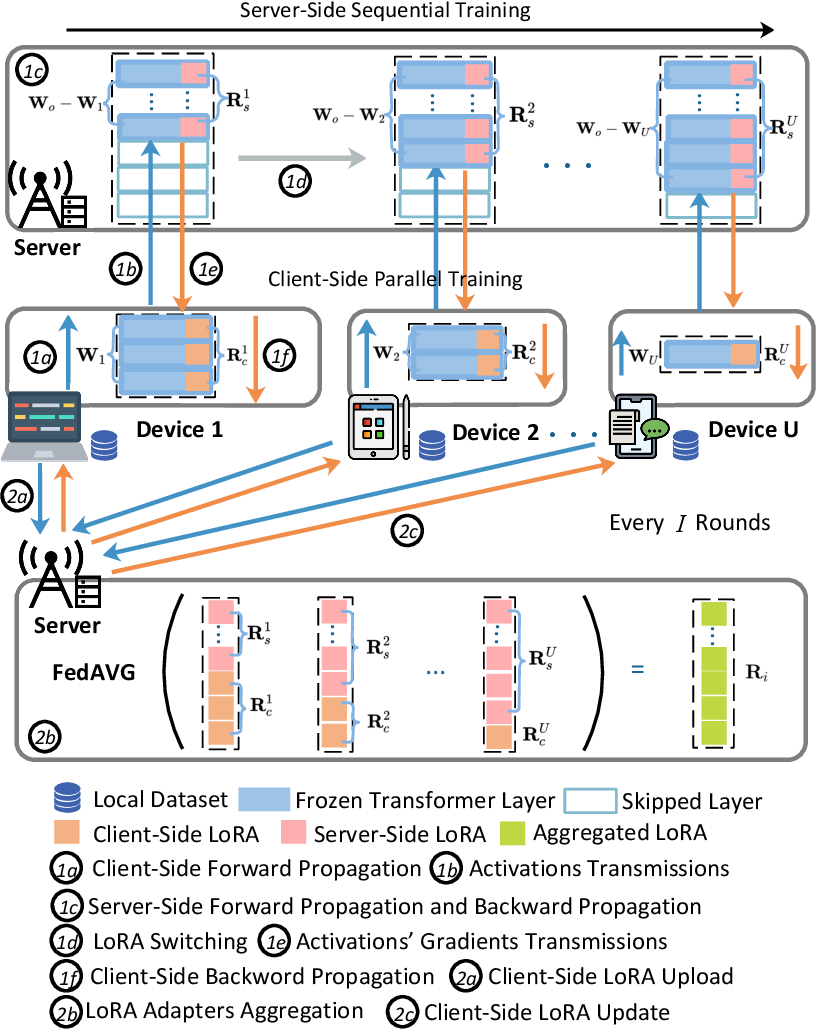}\\
	\caption{An illustration of the memory-efficient SFL framework.}\label{fig.workflow}
\end{figure}

LoRA freezes the original model but inserts a few small modules into different locations. Specifically,  two decomposed low-rank matrices are used to represent the update of the target module: 
\begin{equation}
    \mathbf{W}' = \mathbf{W} + \Delta\mathbf{W} = \mathbf{W} + \mathbf{B}\mathbf{A},
\end{equation}
where $\mathbf{W} \in \mathbb{R}^{m \times m}$ and  $\mathbf{W}' \in \mathbb{R}^{m \times m} $ represent the pre-trained of target modules and the fine-tuned parameters, respectively.  $\mathbf{A} \in \mathbb{R}^{r \times m}$  and $\mathbf{B} \in \mathbb{R}^{m \times r}$ are the low-rank 
decomposition matrices of $\Delta\mathbf{W}$. $m$ is the feature dimension of the hidden layer. $r$ is the LoRA rank. $r << m$ leads to dramatic parameter reduction of $\Delta\mathbf{W}$. LoRA fine-tuning method updates the matrices $\mathbf{A}$ and $\mathbf{B}$ rather than $\mathbf{W}$ directly, resulting in significant savings in GPU memory usage. For ease of presentation, we define the LoRA adapter as $\mathbf{R} = \{\mathbf{A}, \mathbf{B} \}$.   

Although LoRA can reduce the number of training parameters, the memory requirements for models and activations will be a bottleneck for mobile devices. To fit the limited computing and memory resources, each client is allocated a reasonable client-side pre-trained model $\mathbf{W}_u$, and the corresponding client-side LoRA adapters $\mathbf{R}_c^u$.  The computing capability of client $u$ is  $C_u$.
Compared to the clients, the edge server owns powerful computing and memory resources. The server maintains an entire pre-trained model denoted by $\mathbf{W}_o$ and multiple server-side LoRA adapters denoted by $\{\mathbf{R}_s^1,\mathbf{R}_s^2,...,\mathbf{R}_s^u,...,\mathbf{R}_s^U \}$. The server-side LoRA adapters correspond to the client-side LoRA adapters, i.e., the full LoRA adapters of client $u$ are $\mathbf{R}_f^u =  \mathbf{R}_c^u + \mathbf{R}_s^u$.
The server is responsible for the forward propagation and backward propagation computation of the pre-trained models that the clients have left behind. Besides, the server takes charge of synchronizing the LoRA adapters, periodically aggregating the client-side LoRA adapters and the server-side LoRA adapters.  

The goal is to learn an optimal LoRA adapter model that minimizes the global loss function across all the clients: 
\begin{align} 
    \mathbf{R}_f^* &= \arg\min_{\mathbf{R}_f} F(\mathbf{W}_o,\mathbf{R}_f;\{D_u\}  ) \\ \nonumber
    &= \frac{1}{U} \sum_{u=1}^{U}f_u(\mathbf{W}_u,\mathbf{R}_c^u; \mathbf{W}_o-\mathbf{W}_u,\mathbf{R}_s^u;  \{D_u\}),
\end{align}
where $f_u(\cdot) $ is the local loss function of mobile device $u$.

\section{Memory-Efficient SFL Framework}
The proposed framework integrates SL's strengths in model splitting and FL in parallel training and further advances by significantly reducing the server's memory and computational demands. The memory efficiency of the proposed framework is accomplished by reusing pre-trained models to train the corresponding server-side LoRA adapters sequentially. 
\normalem
\begin{algorithm2e}[t]
	\footnotesize
	{\setstretch{1.2}
		\Indentp{-1em}
		\caption{  Memory-Efficient SFL Framework}
		\Indentp{1em}
		\Indentp{-1em}
		\KwIn{ $\mathbf{W}_o$, \{$\mathbf{W}_1$, $\mathbf{W}_2$,...,$\mathbf{W}_u$,..., $\mathbf{W}_U$\},  \{$\mathbf{R}^1_c$, $\mathbf{R}^2_c$,...,$\mathbf{R}^u_c$,..., $\mathbf{R}^U_c$\}, \{$\mathbf{R}^1_s$, $\mathbf{R}^2_s$,...,$\mathbf{R}^u_s$,..., $\mathbf{R}^U_s$\}, $I$,  $D$ }
		\KwOut{ $\mathbf{R}_f^*$  ;}
		\Indentp{1em}
		\SetKwFor{While}{while}{do}{end~while}
		\For{$t = 1,2,...,T$}{  /** On mobile devices **/ \\ \For{$u \in \mathcal{U}$ in parallel }{ $\mathbf{v}_u^t = f(\mathbf{W}_u,\mathbf{R}_c^{u,t-1}; x_u^t);$ \\Send $(\mathbf{v}_u^t, y_u^t)$ to the server;\\ }  /** On the server **/ \\\For{$u \in \mathcal{U}$ in sequential }{ $\hat{y}_u^t = f(\mathbf{W}_o-\mathbf{W}_u,\mathbf{R}_s^{u,t-1};\mathbf{v}_u^t)$; \\  Based on $\hat{y}_u^t$ and $y_u^t$, calculate the loss and then derive the gradients to update the server-side LoRA adapters $\mathbf{R}_s^{u,t}$; \\ Send activations’ gradients to corresponding clients;}  
        /** On mobile devices **/ \\ \For{$u \in \mathcal{U}$ in parallel }{  Based on the received activations’ gradients, update the client-side LoRA adapters $\mathbf{R}_c^{u,t}$  ;\\ } 
          \If { $t$ mod $I$ = 0 } { /** On mobile devices **/ \\  \For{$u \in \mathcal{U}$ in parallel } { Send the client-side LoRA adapters $\mathbf{R}_c^{u,t}$ to the server; }
          /** On the server **/ \\
           \For{$u \in \mathcal{U}$ in parallel }{$\mathbf{R}^u_f = \{\mathbf{R}_c^u,\mathbf{R}_s^u\}$ }
           $\mathbf{A}_{n,i}=\sum_{u=1}^{U}\frac{|D_u|}{|D|}\mathbf{A}^u_{n,i}$;\\ $\mathbf{B}_{n,i}=\sum_{u=1}^{U}\frac{|D_u|}{|D|}\mathbf{B}^u_{n,i}$; \\
           According to Eq. (9), split the aggregated LoRA adapters; \\
           Update the client-side LoRA and the server-side LoRA; } }
           \par}
\end{algorithm2e}

Before the training, each client uploads the resource information (e.g., memory, computing capacity, etc.) of its own device to the server. The server replicates a reasonable client-side submodel for each client based on the provided information and records the model cut points for each client. After completing the training configuration, the training procedure is started until the model converges. As shown in Fig. 1,  the training procedure can be divided into parallel-sequential fine-tuning and LoRA adapter aggregation two phases, which are detailed below. 
\subsubsection{Parallel-Sequential Fine-Tuning} 
The parallel-sequential fine-tuning phase involves parallel fine-tuning on the clients and sequential fine-tuning on the server, which includes the following steps:

\begin{itemize}
  \item Client-side model forward propagation: Each client performs forward propagation computation of the client-side pre-trained model locally. In particular, client $u$ randomly samples a mini-batch $\{x_u^t,y_u^t\} \subseteq  D_u $ in training round $t$, where $x_u^t$ and $y_u^t$ represent the input sequences and the corresponding label, respectively. The batch size is $B$. After processing a mini-batch of data through the client-side model, the activations are generated at the split layer. In training round $t$, the activations of client $u$ can be given by
	\begin{equation}
		\mathbf{v}_u^t = f(\mathbf{W}_u,\mathbf{R}_c^{u,t-1}; x_u^t),
	\end{equation}
   where $\mathbf{R}_c^{u,t-1}$ is the LoRA adapters of client $u$ updated in training round $t-1$. 
   
\item Activations transmissions: After the client completes the forward propagation, the activation, the corresponding label, and the corresponding index of the split layer are uploaded to the server via wireless communication. The server will utilize the activation for subsequent fine-tuning. Unlike typical federated learning, each client can upload the activation on its own terms and does not need to synchronize the uploads.
	
\item Server-side model forward propagation and backward propagation: Based on the received client information, the corresponding server-side LoRA adapters are loaded, and the received activations are passed into the corresponding cut layer for forward propagation computation in the remaining model. Then, the predicted value can be given as 
	\begin{equation}
		\hat{y}_u^t = f(\mathbf{W}_o-\mathbf{W}_u,\mathbf{R}_s^{u,t-1};\mathbf{v}_u^t).
	\end{equation}
The predicted value and the label are employed to compute the loss and then derive the gradients of the server-side LoRA adapters. Moreover, the server-side LoRA adapters can be updated by backward propagation.
   
\item LoRA adapters switching: After completing the update of the server-side LoRA adapters of the current client, the server needs to update the server-side LoRA adapters of the next client.  Based on the information received from the client, the server switches the current LoRA adapters and loads the next training LoRA adapters.
	
\item Activations’ gradients transmissions: After the server-side backward propagation is completed, the server transmits the activations'  gradients to its corresponding client. This step is performed simultaneously with 1d. 
	
\item Client-side model backward propagation: The client fine-tunes its client-side pre-trained model based on the received activations’ gradients.
\end{itemize}

\subsubsection{LoRA Adapter Aggregation} The LoRA adapter aggregation phase focuses on aggregating all clients' full LoRA adapters on the server and then updating the server-side LoRA adapters and client-side LoRA adapters, which is executed per $I$  training round. This phase includes the following three steps.   
\begin{itemize}
\item Client-side LoRA adapters upload: Each client sends its client-side LoRA adapters to the server via wireless communications. The client-side LoRA adapters and the corresponding server-side LoRA adapters combine to form a full LoRA adapters for aggregating. The full LoRA adapters of client $u$ can be given by 
 \begin{equation}
	\mathbf{R}^u_f = \{\mathbf{R}_c^u,\mathbf{R}_s^u\}=\{\mathbf{A}^u_1,\mathbf{B}^u_1,\mathbf{A}^u_2,\mathbf{B}^u_2,...,\mathbf{A}^u_N,\mathbf{B}^u_N\},
	\label{lora} 
	\end{equation}
    where $N$ is the total number of trainable LoRA adapters of pre-trained model.
 \item Full LoRA adapters aggregation: The server aggregates all clients' full LoRA adapters into aggregated LoRA adapters. In aggregation round $i$, the decomposition matrices $\mathbf{A}$ and $\mathbf{B}$ of $n$-th LoRA adapter are aggregated separately as follows
	\begin{equation}
		\mathbf{A}_{n,i}=\sum_{u=1}^{U}\frac{|D_u|}{|D|}\mathbf{A}^u_{n,i},
	\end{equation}
    \begin{equation}
	\mathbf{B}_{n,i}=\sum_{u=1}^{U}\frac{|D_u|}{|D|}\mathbf{B}^u_{n,i}.
    \end{equation}
    Then, in aggregation round $i$,  we can obtain the aggregated full LoRA adapters   
    \begin{equation} 
    	\mathbf{R}_i = \{\mathbf{A}_{1,i},\mathbf{B}_{1,i},\mathbf{A}_{2,i},\mathbf{B}_{2,i},...,\mathbf{A}_{N,i},\mathbf{B}_{N,i}\}.
     \end{equation}
 
\item Client-side LoRA adapters update:  After completing full LoRA adapters aggregation, the server sends the aggregated client-side LoRA adapters to each client. Specifically, the server splits the full LoRA adapters to fit the client-side pre-trained model. For client $u$, the process can be represented as follows:
	 \begin{align} 
		\mathbf{R}_i =& \{\underbrace{\mathbf{A}_{1,i},\mathbf{B}_{1,i},...,\mathbf{A}_{N_c^u,i},\mathbf{B}_{N_c^u,i}}_{\mathbf{R}_c^{u,i}}, \nonumber \\ &\underbrace{\mathbf{A}_{N_c^u+1,i},\mathbf{B}_{N_c^u+1,i},...,\mathbf{A}_{N,i},\mathbf{B}_{N,i}}_{\mathbf{R}_s^{u,i}} \},
	\end{align}
where $N_c^u$ is the number of client-side LoRA adapters of client $u$. Correspondingly, $N-N_c^u$ is the number of server-side LoRA adapters of client $u$. $\mathbf{R}_c^{u,i}$ and $\mathbf{R}_s^{u,i}$ are the aggregated client-side LoRA adapters and server-side LoRA adapters in aggregation round $i$, respectively. Afterward, each client utilizes the received aggregated LoRA adapters as the initial LoRA configuration for the next training round.  
\end{itemize}

As shown in Fig. 1, steps 1a, 1b, and 1f are performed in parallel on the clients to improve time efficiency. Steps 1c, and 1e are executed sequentially on the server to reduce the pressure on memory and computation. Step 1d is a bridge from parallel to sequential. The details of the proposed memory-efficient SFL framework are presented in Algorithm 1.

\section{Processing Sequence Scheduling} 
In the proposed framework, client-side submodels are trained in parallel at the clients, while server-side submodels are trained sequentially on the server. The order of sequential computation impacts the backward propagation of the clients. The decision of client order for server-side training influences the overall training time. Therefore, this section focuses on the training order scheduling.  

For client $u$,  the training time of one step for client $u$ can be expressed as 
\begin{equation}
	T_u =  T_{u}^f + T_u^{fc} + T_{u}^{w} + T_{u}^s + T_u^{bc} + T_{u}^b,
\end{equation}
where $T_{u}^f$ is the time of client-side model forward propagation. $T_{u}^s$ is the time of server-side computation, which includes server-side model forward propagation and backward propagation. $T_{u}^b$ is the time of client-side model backward propagation. $T_u^{fc}$ and $T_u^{bc}$  are the activations transmission time and the gradients transmission time, respectively.  $T_{u}^{w}$  is the waiting time since the server-side LoRA adapters are updated sequentially by the server.

Assuming the order of client $u$ is $p_u$, then the waiting time $T_{u}^{w}$ can be expressed as:

\begin{equation}
	T_{u}^{w} =  \sum_{i \in \mathcal{U}_{p_u} } T_{i}^s, 
\end{equation}
where $\mathcal{U}_{p_u}$ is the set of clients whose orders are less than $p_u$.

In the proposed framework, the completion of one training step can be expressed as:
\begin{equation}
	\max_{u \in \mathcal{U}}\ T_u.
\end{equation}

\normalem

\begin{algorithm2e}[t]
	\footnotesize
	{\setstretch{1.2}
		\Indentp{-1em}
		\caption{ Training Order Scheduling Algorithm}
		\Indentp{1em}
		\Indentp{-1em}
		\KwIn{ Computing capability of the clients \{$C_1$, $C_2$,...,$C_u$,..., $C_U$\};   Number of client-side LoRA adapters of the clients \{$N_c^1$, $N_c^2$,..., $N_c^u$,..., $N_c^U$\};}
		\KwOut{Training order $\mathcal{P}$;}
		\Indentp{1em}
		\SetKwFor{While}{while}{do}{end~while}
		\For{$\forall u \in {\cal U}$}{
			Calculate ${N_c^u}/{C_u}$;\\ }
		Sort $u \in  {\cal U}$ according to ${N_c^u}/{C_u}$ in descending order.;\\
		Assign the order according to the index of the sorted collection;\\
		The training order $\mathcal{P}$ is obtained based on the sorting results;
		\par}
\end{algorithm2e}

To minimize the training time of each step, we need to schedule the training order. The optimization problem can be expressed as:
\begin{align}
  \min_{\mathcal{P}} &\max_{u \in \mathcal{U}}\ T_u \\
  \text{s.t.} & \ p_i \neq p_j, \forall p_i, p_j \in \mathcal{P}, \forall i, j \in \mathcal{U}\\
  &\ p_u \in \{1,2,...,U\} , \forall u \in \mathcal{U} 
\end{align}
where constraint (14) represents that the order of each client is unique. Constraint (15) specifies the range of values for the order, which corresponds to the number of the clients. We can observe that this problem is a pipeline problem. To minimize the overall completion time, the key is to hide the communication time and the client computation time under the server computation time as much as possible \cite{9155272}. Since gradient size passed through each layer is the same and the gradients transmission time is much smaller than the backward propagation time.    Therefore, we use a greedy-based algorithm where the server prioritizes tasks with a longer backward propagation time of the client (i.e. $T_{u}^b$).  

The client-side backward propagation time $T_{u}^b$ depends on the client-side LoRA $\mathbf{R}_c^u$ and the computing capacity $C_u$. Therefore, we use the number of client-side LoRA adapters of the clients and the client's computing capacity to give an order to the client backward propagation time. The above training order scheduling method is summarized in Algorithm 2. 
\begin{table*}[t]
	\centering
	\caption{Performance Comparison of Different Schemes}
	\label{tab:comparison}
	\begin{tabular}{l|c|c|c|c|c}
		\hline
		\textbf{Schemes} & \textbf{Memory Consumption (MB)} & \textbf{Convergence Round} & \textbf{Convergence Time (s)} & \textbf{Accuracy} & \textbf{F1} \\ \hline
		SL           & 1346.85                    & 89                            & 57341.78
		& 0.8925                     &0.8948                     \\ \hline
		SFL      & 7327.90                    & 180                           & 35654.90                      & 0.8935                     & 0.8937                     \\ \hline
		Ours   & 1482.63                    & 180                           & 33471.70                      & 0.8935                     & 0.8937                    \\ \hline
	\end{tabular}
	\begin{flushleft}
		\footnotesize{*Memory consumption is measured on the server.}
	\end{flushleft}
\end{table*}
\begin{figure*}[htbp]
	\centering
	\begin{subfigure}{0.32\textwidth}
		\includegraphics[width=\textwidth]{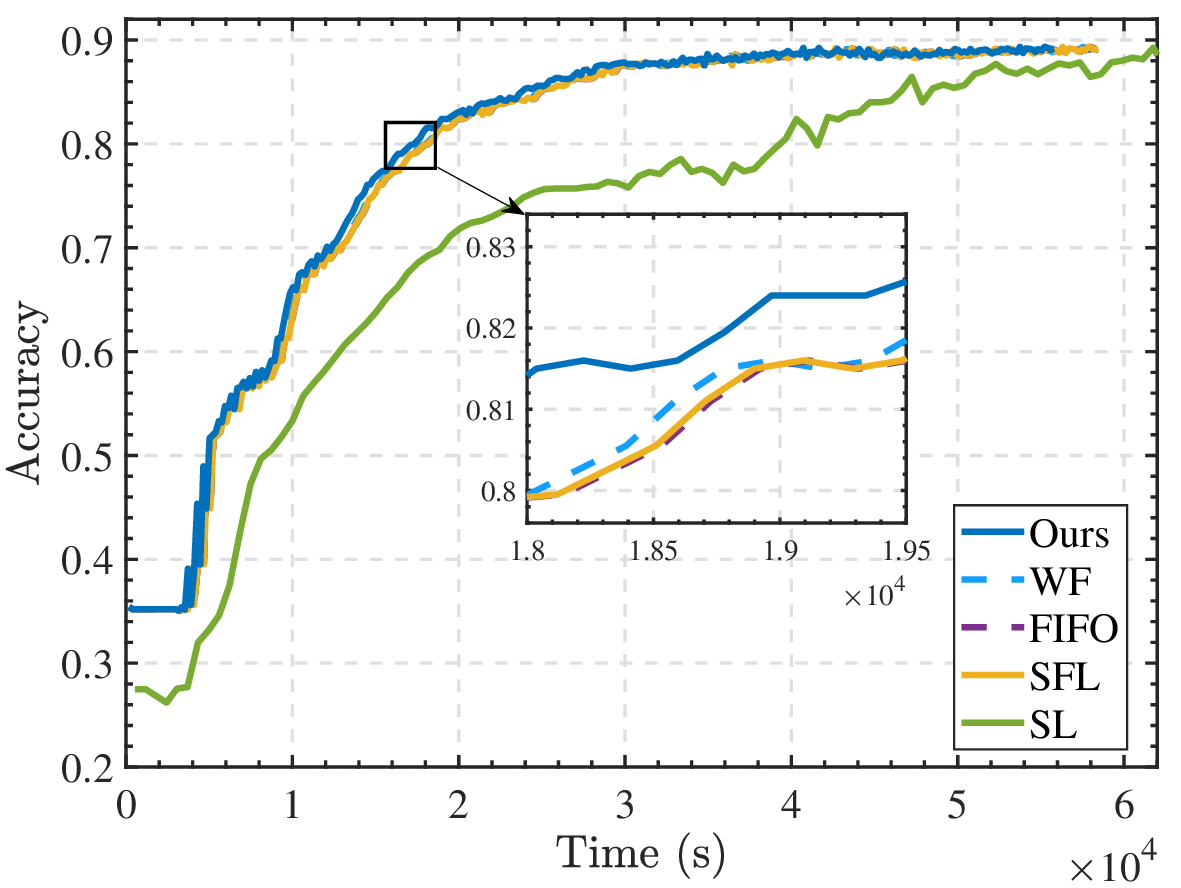} 
		\caption{Training performance (accuracy) vs. time.}  
		\label{fig:subfig1}
	\end{subfigure}
	\hfill
	\begin{subfigure}{0.32\textwidth}
		\includegraphics[width=\textwidth]{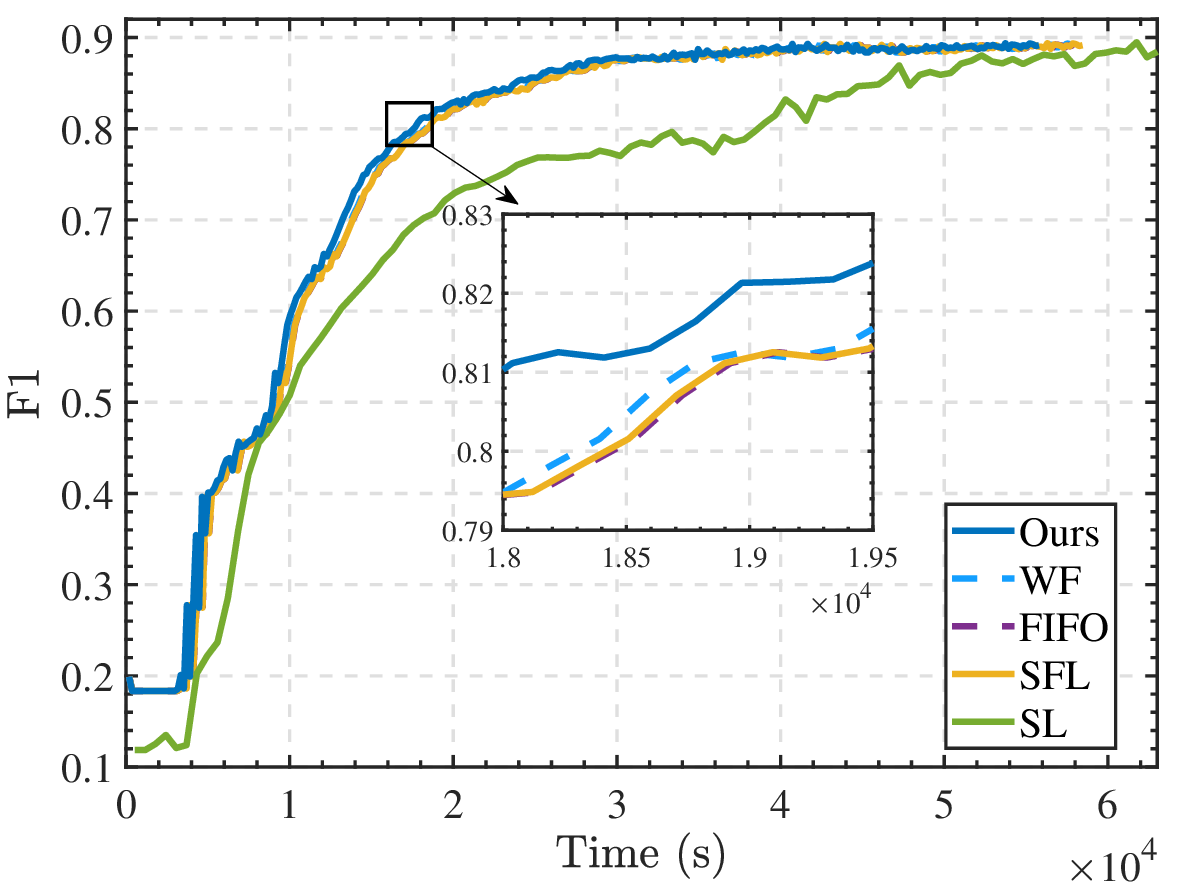} 
		\caption{Training performance (f1) vs. time.}  
		\label{fig:subfig2}
	\end{subfigure}
    \hfill
	\begin{subfigure}{0.32\textwidth}
		\includegraphics[width=\textwidth]{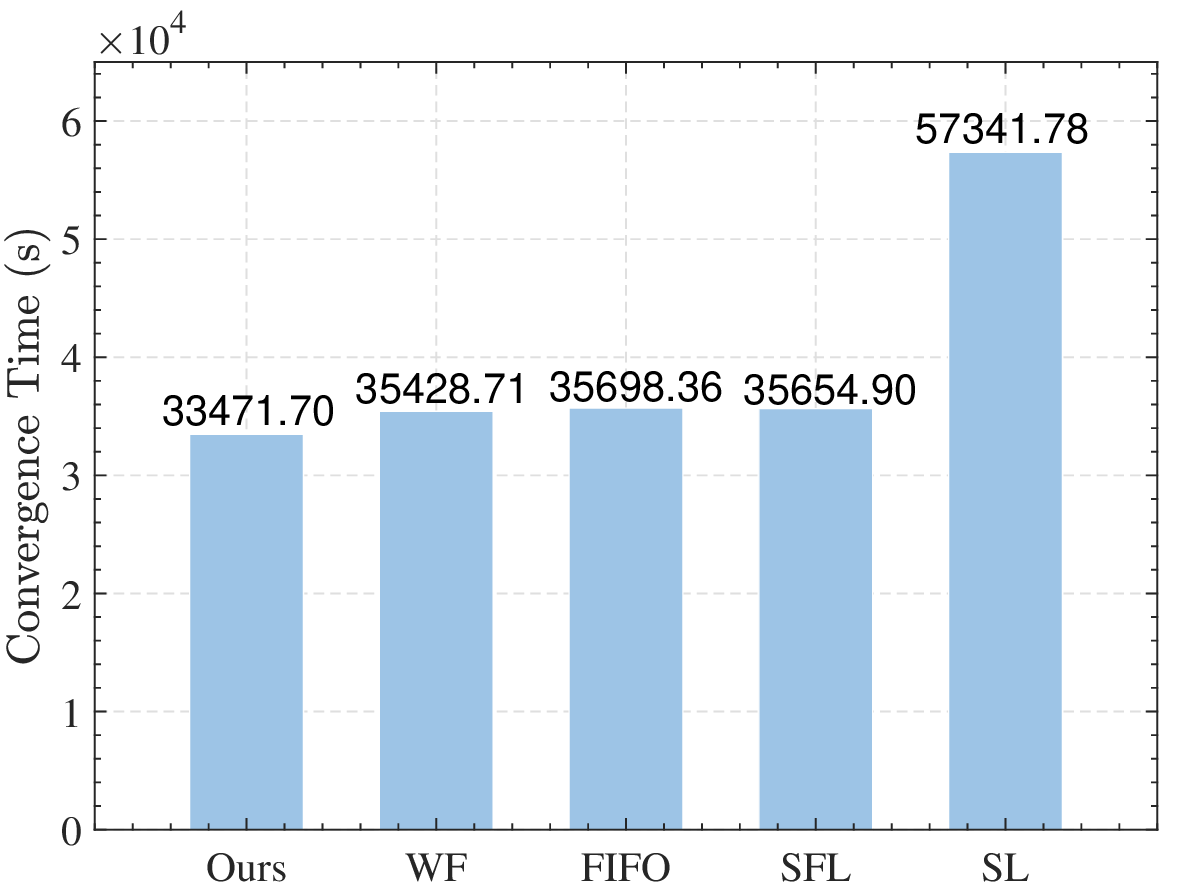} 
		\caption{Convergence time of different schemes.}  
		\label{fig:subfig2}
	\end{subfigure}
	\caption{Training performance under different schemes.}
	\label{fig:overall}
\end{figure*}
\section{Performance Evaluation}
\subsection{Simulation Setup}
The performance of the proposed scheme is evaluated in this section.  We leverage BERT-base \cite{devlin2018bert} as the pre-trained model for text analysis tasks using CARER dataset \cite{saravia-etal-2018-carer}.  We use an RTX 4080s server with a computational capability of 52.2 TFLOPS and consider six heterogeneous clients: a Jetson Nano (0.472 TFLOPS) with first one transformer layer, a Jetson TX2 (1.33 TFLOPS) with first one transformer layer, a Snapdragon 8s Gen 3 (1.689 TFLOPS) with first two transformer layers, a Snapdragon 8 Gen 3 (2.774 TFLOPS) with first two transformer layers, an A17 Pro (2.147 TFLOPS) with first three transformer layers, and an M3 (3.533 TFLOPS) with first three transformer layers. The data rate between the server and each client is set to 100 Mbps. We set the LoRA rank, batch size, learning rate, and maximum sequence length to 16, 16, 0.00001, and 128. 

To investigate the advantages of the proposed scheme, we compare the proposed scheme with the following baselines.
\begin{itemize}
    \item \textbf{Split Learning (SL)} \cite{10040976}: The model is split into a client-side submodel and a server-side submodel. After completing the training of a client, the client uploads the client's model to the server to update the overall model. After that, the server splits the model and sends it to the next client to continue training. 
    \item \textbf{Split Federated Learning (SFL)} \cite{tian2022fedbert}: The server maintains multiple server-side models corresponding to the multiple heterogeneous clients to support the parallel training at the server. After completing the training, the clients upload the client-side model and the global model is updated by aggregating it on the server.
    \item \textbf{First-in-First-out Scheduling (FIFO)} \cite{9900235}: In the proposed framework, the training order of the server-side submodel is determined by the first-in-first-out approach.   
    \item \textbf{Workload-First Scheduling (WF)} \cite{10400885}: In the proposed framework, the training order of the server-side submodel depends on the required server workload for the task. The task with a large workload is trained first.
\end{itemize}

\subsection{Simulation Results}
Table I shows the memory usage, the convergence round, convergence time, accuracy, and f1 of the three fine-tuning schemes. It can be seen that the three schemes can achieve comparable performance. Although SL has the smallest memory usage, it results in the longest convergence time. SFL takes up a very large amount of memory due to the need to maintain multiple large models. The proposed scheme is able to achieve the fastest convergence. Compared with SL, the proposed scheme reduces the training time by 40\% at the 10\% memory cost. Compared to SFL, the proposed scheme reduces 79\% of memory and 6\% of training time. The proposed scheme reduces the memory footprint by maintaining a full LLM at the server to reuse sequential training. On the other hand, since the multi-model parallel computation of SFL leads to fragmentation of server computational resources, it increases the competition for memory access during computation. The proposed scheme rationally utilizes the server's computational resources and then reduces the overall training latency by designing a reasonable training scheduling scheme.

Figures 2(a) and 2(b) illustrate the performance metrics over training time for accuracy and f1 score, respectively. We can observe that at first, the training performance is stagnant, then grows rapidly, and finally grows slowly to reach convergence. The effect of SL fluctuates because the clients' local datasets are non-IID. Meanwhile, the proposed scheme and SFL improve the performance of model training by integrating data from different distributions through model aggregation, which helps the model learn a wider range of features and patterns. It can be observed from the enlarged sub-graphs in Fig. 2(a) and Fig. 2(b) that the proposed scheme outperforms the FIFO scheduling, the WF scheduling, and the SFL. This shows that our proposed scheduling method can speed up the training time and achieve better model performance in the same training time.

Figure 2(c) presents the convergence time of different schemes. It can be observed that our scheme has the minimum convergence time. Compared to SL and SFL, the proposed scheme can reduce the training time by 41\% and 6.1\%, respectively. The reason is that the proposed scheme improves computational efficiency by computing in parallel with the clients and avoids competition for computing resources by computing sequentially on the server, which effectively utilizes the powerful computational resources of the server. Under the proposed training framework, the proposed scheduling scheme is able to reduce the training time by 5.5\% and 6.2\% compared to WF and FIFO, respectively.   The proposed scheduling method can hide the communication time and client computation time under the server computation time to reduce the overall training time.

\section{Conclusion}
In this paper, we have presented a memory-efficient SFL training framework for LLM fine-tuning over heterogeneous devices. The core of the proposed framework lies in the reuse of the entire LLM by skipping over heterogeneous client-side submodels, which avoids the storage footprint of multiple heterogeneous server-side submodels and the memory footprint of parallel training of multiple server-side submodels. Client-side parallel training improves the training efficiency while server-side sequential training significantly reduces memory pressure and efficiently utilizes computational resources. Furthermore, we consider the training sequence of the server side to minimize the training time. Our experiments have shown that, compared to the baselines, the proposed scheme can reduce memory footprint and fine-tuning delay, demonstrating the potential for LLM-based edge intelligence.

\normalem
\footnotesize
\bibliographystyle{IEEEtran}
\bibliography{IEEEabrv,references}


\end{document}